\documentclass{article}[12pt]

\usepackage{amsmath}
\usepackage{amsmath,amssymb,amsthm,graphicx}
\usepackage{amsfonts}
\usepackage{pstricks}
\usepackage{verbatim}
\usepackage{bbm}
\usepackage{mathrsfs, hyphenat}
\usepackage[makeroom]{cancel}
\usepackage[hidelinks]{hyperref}  
\usepackage[anythingbreaks]{breakurl}  
\usepackage[all]{nowidow} 
\usepackage{amsthm}
\usepackage{comment}
\def\spaceoftensors{{\mathcal T}^3}
\def\spaceoftensorsq{{\mathcal T}^3_q}
\def\spaceoftensorsbarq{{\mathcal T}^3_{\bar q}}
\def\su5{\mbox{SU}(5)}
\def\unitary{\mbox{U}(5)}

\newtheorem{thm}{Theorem}[section]
\newtheorem{defn}[thm]{Definition}

\newtheorem{prop}[thm]{Proposition}
\title{$\mbox{SO}(3)$-Irreducible Geometry in Complex Dimension Five and Ternary Generalization of Pauli Exclusion Principle}

\author{Viktor Abramov\\
{\small Institute of Mathematics and Statistics}\\ {\small University of Tartu, Estonia}
    \and Olga Liivapuu\\
         {\small Institute of Forestry and Engineering}\\ 
          {\small Estonian Univesity of Life Sciences, Estonia}}
\date{}

\begin{document}
\maketitle
\begin{abstract}{We propose a notion of a ternary skew-symmetric covariant tensor of 3rd order, consider it as a 3-dimensional matrix and study a ten-dimensional complex space of these tensors. We split this space into a direct sum of two five-dimensional subspaces and in each subspace there is an irreducible representation of the rotation group $\mbox{SO}(3)\hookrightarrow\mbox{SO}(5)$. We find two independent
$\mbox{SO}(3)$-invariants of ternary skew-symmetric tensors, where one of them is the Hermitian metric and the other is the quadratic form. We find the stabilizer of this quadratic form and its invariant properties. Making use of these invariant properties we define a $\mbox{SO}(3)$-irreducible geometric structure on a five-dimensional complex Hermitian  manifold. We study a connection on a five-dimensional complex Hermitian  manifold with a $\mbox{SO}(3)$-irreducible geometric structure, find its curvature and torsion. The structures proposed in this paper and their study are motivated by a ternary generalization of the Pauli's principle proposed by R. Kerner.}
\end{abstract}
\section{Introduction}
The concept of skew symmetry underlies many structures of modern algebra and geometry. The following structures can serve as examples of such structures: Lie algebras, Grassmann algebras, algebras of differential forms on a manifold, Clifford algebras. In the case of a Lie algebra its Lie bracket is skew-symmetric with respect to a permutation of arguments of this bracket. A Grassmann algebra can be constructed by means of covariant skew-symmetric tensors of some finite dimensional vector space. In this case the skew symmetry of a tensor means that the rearrangement of any two subscripts leads to the change of the sign of a tensor, and the exterior multiplication of two such tensors is defined with the help of the alternation of the tensor product of these tensors. At the end of the last century and the beginning of this century, we witnessed the emergence of interest towards an $n$-ary generalization of Lie algebra, that is, a generalization in which a Lie bracket contains $n$ arguments. The concept of skew symmetry can be easily extended to $n$-ary multiplications if we require that any inversion of two arguments in an $n$-ary product results in a change in sign. An $n$-ary bracket of $n$-Lie algebra is skew-symmetric precisely in this classical sense. However, when moving from a binary law of multiplication to an n-ary one, where n>2, we can pose an interesting question of possible analogues of the classical concept of skew symmetry. In order to formulate the concept of skew symmetry of an algebraic operation we use permutations of factors in the product. In the case of binary multiplication we have only one permutation of variables in this multiplication and, consequently, we have only one notion of skew-symmetric binary product. By requiring that an $n$-ary product of elements ($n\geq 2$) be equal to zero whenever any two elements in this product are equal, we obtain an equivalent form of the concept of skew symmetry (an $n$-ary product is assumed to be linear in each argument). This formulation of skew symmetry explains why the concept of skew symmetry plays an important role in theoretical physics. Recall the Pauli exclusion principle, which states that two fermions in a quantum system cannot co-exist if they have identically equal sets of quantum numbers. It follows then that a wave function of a quantum system containing identical sets of quantum numbers of two fermions must vanish. Now the skew-symmetry of a wave function with respect to permutation of quantum states of any two fermions follows from the linearity of a wave function.

If we consider a ternary multiplication then in a ternary product we
have six possible permutations of arguments, where three of them are inversions (non-cyclic permutations), and three are cyclic.It is natural to use non-cyclic permutations to extend binary skew-symmetry to ternary multiplication laws, that is, we call a ternary multiplication totally skew-symmetric if it is skew-symmetric with respect to any pair of arguments. It is precisely this notion of skew symmetry that is used in 3-Lie algebras, that is, any non-cyclic permutation of arguments in a ternary Lie bracket of 3-Lie algebra changes a sign of this bracket. Equivalently, if among the three elements of a ternary Lie bracket there are two equal ones, then regardless of where in the bracket these equal elements appear, the ternary bracket is equal to zero. In this formulation we see a direct connection with the classical Pauli exclusion principle.

But in the case of ternary multiplication, unlike the binary one, we have three more cyclic permutations, and these permutations can be used to construct a ternary analogue of skew symmetry, different from the one indicated above. Let $V$ be a finite dimensional vector space with a skew-symmetric binary law of multiplication $(u,v)\in V\times V\to u\cdot v\in V$. Then the skew-symmetry can be expressed in two equivalent ways
\begin{eqnarray}
&& u\cdot v+v\cdot u=0,\label{introduction:skewsymmetry 1}\\
&& u\cdot v=-v\cdot u.\label{introduction:skewsymmetry 2}
\end{eqnarray}
Let $(W,\tau)$ be a complex finite dimensional vector space equipped with a ternary trilinear law of multiplication
$$
\tau: (x,y,z)\in W\times W\times W\to \tau(x,y,z)\in W.
$$
Since we use cyclic permutations a ternary analog of (\ref{introduction:skewsymmetry 1}) is
\begin{equation}
\tau(x,y,z)+\tau(y,z,x)+\tau(z,x,y)=0.
\label{introduction:cyclic relation}
\end{equation}
It is natural to raise the question of what could serve as a ternary analogue of relation (\ref{introduction:skewsymmetry 2}).
The answer to this question is the relations
\begin{equation}
\tau(x,y,z)=q\;\tau(y,z,x)=\bar q\;\tau(z,x,y),
\label{introduction:q-relations}
\end{equation}
or their conjugate version
\begin{equation}
\tau(x,y,z)=\bar q\;\tau(y,z,x)=q\;\tau(z,x,y)
\label{introduction:barq-relations}
\end{equation}
where $q=\exp(2i\pi/3)$ is the primitive 3rd order root of unity. It is important to note that there is a significant difference between binary relations (\ref{introduction:skewsymmetry 1}),
(\ref{introduction:skewsymmetry 2}) and ternary relations (\ref{introduction:cyclic relation}),(\ref{introduction:q-relations}). In the binary case, the relations (\ref{introduction:skewsymmetry 1}) and
(\ref{introduction:skewsymmetry 2}) are equivalent, one immediately follows from the other. This is not true in the case of a ternary law of multiplication. Relations (\ref{introduction:cyclic relation}) and (\ref{introduction:q-relations}) are not equivalent. Due to the well-known property of cubic roots of unity $1+q+\bar q=0$, the relations (\ref{introduction:cyclic relation}) follow from the relations (\ref{introduction:q-relations}), but not vice-versa. By other words (\ref{introduction:cyclic relation}) is a more general condition than (\ref{introduction:q-relations}).

Thus, an analogue of the notion of skew symmetry in the case of ternary multiplication can be one of the conditions (\ref{introduction:cyclic relation}), (\ref{introduction:q-relations}), (\ref{introduction:barq-relations}) with (\ref{introduction:q-relations}), (\ref{introduction:barq-relations}) being particular cases of (\ref{introduction:cyclic relation}). But in any case, no matter which of the conditions (\ref{introduction:cyclic relation}), (\ref{introduction:q-relations}), (\ref{introduction:barq-relations}) we take as a ternary analogue of the concept of skew symmetry, for any $x\in W$ we will have $\tau(x,x,x)=0$, and generally $\tau(x,x,y)\neq 0$, where $x\neq y$. Consequently, a ternary analogue of skew symmetry defined either by (\ref{introduction:cyclic relation}) or (\ref{introduction:q-relations}) or (\ref{introduction:barq-relations}) is significantly different from the classical concept of skew symmetry in the sense that in the case of classical skew symmetry, the presence of two equal elements in a product leads to zero, and in the case of ternary skew symmetry defined by one of the conditions (\ref{introduction:cyclic relation}), (\ref{introduction:q-relations}), (\ref{introduction:barq-relations}) the presence of two equal elements in a ternary product in general does not imply that this product is equal to zero. But at the same time, if all three elements in a ternary product are equal, that is $x=y=z$, then the product is equal to zero $\tau(x,x,x)=0$.

An analogue of the Pauli exclusion principle, which is based on a ternary skew symmetry, defined by one of conditions (\ref{introduction:cyclic relation})-(\ref{introduction:barq-relations}), could be formulated as follows: Three particles cannot coexist in a quantum system if these three particles are in the same quantum state, but two such particles can. It is in this form that an analogue of the Pauli exclusion principle was proposed by Richard Kerner, who calls it a ternary generalization of the Pauli exclusion principle. Richard Kerner argues that the ternary generalization of the Pauli exclusion principle applies to quarks. In the quark model, quarks are considered as fermions, and three quarks or three anti-quarks form a baryon. The ternary generalization of the Pauli exclusion principle can in this case be formulated as: Three quarks in the same quantum state cannot form a stable configuration, observed as one of strongly interacting particles, but at the same time, the coexistence of two quarks with the same isospin value is possible.

Relations (\ref{introduction:cyclic relation})-(\ref{introduction:barq-relations}) can be used to construct an analogue of Grassmann algebra, that is, we can consider an algebra over the field of complex numbers generated by a system of generators that obey one of the relations (\ref{introduction:cyclic relation})-(\ref{introduction:barq-relations}). The properties, structure and possible applications of such algebras were studied in papers \cite{Abramov_Kerner_LeRoy:J_Math_Phys_1997,Abramov_IJGMMP_2023,Kerner:J_Math_Phys_1992}. These algebras can be applied to construct a generalization of  exterior calculus with exterior differential $d$ satisfying $d^3=0$ \cite{Abramov_Kerner:J_Math_Phys_2000}. A generalization of the Dirac operator based on the mentioned above algebras and relation to a ternary generalization of Pauli exclusion principle can be found in \cite{Kerner:Acad_Sci_Paris_1991,Kerner:J_Math_Phys_1992,Kerner:Phys_Atom_Nucl_2017,Kerner:Universe_2019} to In this article, the main object of study is the subspace of an algebra whose generators obey relations (\ref{introduction:cyclic relation}), spanned by the triple products of the generators. This subspace can be identified with the space of complex-valued covariant third-order tensors $T_{ijk}$ in three-dimensional space which have the following property
\begin{equation}
T_{ijk}+T_{jki}+T_{kij}=0.
\label{introduction:cyclic condition}
\end{equation}
If we impose an additional condition on the tensors of this space, which is
\begin{equation}
T_{iij}=0,\;\;T_{iji}=0,\;\;T_{jii}=0,
\label{introduction:trace is zero}
\end{equation}
i.e. the trace of a tensor over any pair of subscripts is equal to zero, we obtain the space of tensors known in the theory of representations of the rotation group \cite{Gelfand-Minlos-Shapiro}. More precisely, the space of tensors satisfying conditions (\ref{introduction:cyclic condition}), (\ref{introduction:trace is zero}) is ten-dimensional and in this space we have a twofold irreducible tensor representation of the rotation group. Let us denote this ten-dimensional complex vector space by $\mathcal T^3$. A twofold irreducible tensor representation of the rotation group in $\mathcal T^3$ splits into two irreducible tensor representations if we decompose the ten-dimensional representation space $\mathcal T^3$ into a direct sum of two five-dimensional subspaces in a way invariant under the action of the rotation group. A decomposition into two subspaces can be made with the help of the relations (\ref{introduction:q-relations}), (\ref{introduction:barq-relations}), that is, we define the subspace $\mathcal T^3_q\subset \mathcal T^3$ by imposing the additional condition
\begin{eqnarray}
T_{ijk}=q\;T_{jki}=\bar q\;T_{kij}.
\label{introduction:q-relations for tensors}
\end{eqnarray}
It is easy to see that in this case the condition (\ref{introduction:cyclic condition}) follows from (\ref{introduction:q-relations for tensors}).
Hence we have
\begin{equation}
{\mathcal T}^3_q=\{T=(T_{ijk}\in{\mathcal T}^3):\;\;T_{ijk}=q\;T_{jki}=\bar q\;T_{kij}\},
\end{equation}
and analogously
\begin{equation}
{\mathcal T}^3_{\bar q}=\{T=(T_{ijk}\in{\mathcal T}^3):\;\;T_{ijk}={\bar q}\;T_{jki}=q\;T_{kij}\}.
\end{equation}
Then ${\mathcal T}^3={\mathcal T}^3_q\oplus {\mathcal T}^3_{\bar q}$ and in each of subspaces ${\mathcal T}^3_{q},{\mathcal T}^3_{\bar q}$ we have an irreducible representation of the rotation group. It is known that every representation of the rotation group can be made unitary if we endow a representation space with an appropriate Hermitian metric. We endow the space $\mathcal T^3_q$ with the Hermitian metric
\begin{equation}
h(T,S)=T_{ijk}\overline{S}_{ijk}
\end{equation}
and show that the irreducible representation of the rotation group in $\mathcal T^3_q$ is an inclusion $R:\mbox{SO}(3)\hookrightarrow\mbox{SU}(5)$. We find the orthonormal basis $E_A, 1\leq A\leq 5$ (here $E_A$ are complex-valued 3rd order covariant tensors satisfying (\ref{introduction:trace is zero}), (\ref{introduction:q-relations for tensors})) for the Hermitian space $\mathcal T^3_q$ and identify the space $\mathbb C^5$ with the Hermitian vector space of tensors $\mathcal T^3_q$ by putting
$$
z=(z^A)\in {\mathbb C}^5\to T(z)=z^A\,E_A\in{\mathcal T}^3_q.
$$
Then the irreducible representation $R$ of the rotation group can be written in the form
\begin{equation}
g_{im}\,g_{jl}\,g_{kp}\;(E_B)_{mlp}=(R(g))^A_B\,(E_A)_{ijk},
\end{equation}
where $g=(g_{ij})\in\mbox{SO}(3))$. We calculate all $\mbox{SO}(3)$-invariants of the representation $R$ and this calculation shows that there are only two non-trivial independent invariants. Obviously one of them is the canonical Hermitian metric $h(z,\bar z)=\sum_A\,z^A\,\bar z^A$ and the other is the quadratic form
\begin{equation}
K(z,z)=K_{AB}\;z^Az^B=(z^1)^2+(z^2)^2+(z^3)^2+2q\,z^4z^5.
\end{equation}
We study the properties of the quadratic form $K(z,z)$. Particularly we show that the matrix $K_{AB}$ of the quadratic form $K(z,z)$ is symmetric, unitary and its determinant is the 6th order primitive root of unity $\epsilon=\exp(\pi\,i/3)$. These properties are invariant under action of the unitary group $\mbox{U}(5)$ in the five-dimensional complex space $\mathcal T^3_q$. Then we find the subgroup of the group $\mbox{SU}(5)$ which is a stabilizer of the quadratic from $K(z,z)$ in the five-dimensional complex vector space $\mathcal T^3_q$. In analogy with approach proposed in \cite{Friedrich} and developed in \cite{Bobienski and Nurowski} we define a $\mbox{SO}(3)$-irreducible geometric structure in complex dimension 5 and study its geometry.
\section{Five-dimensional complex space of $\mbox{SO}(3)$-ir\-re\-ducible representation}
The aim of this section is to describe an irreducible tensor representation of the rotation group. In what follows we consider complex-valued covariant tensors defined in 3-dimensional Euclidean space $\mathbb R^3$. Let $T=(T_{i_1i_2\ldots i_p})$ be a tensor of rank $p$. In what follows, we will use the Einstein convention of summation over repeated indices. Then the formula
\begin{equation}
\Tilde{T}_{j_1j_2\ldots j_p}=g_{i_1j_1}g_{i_2j_2}\ldots g_{i_pj_p}\;T_{i_1i_2\ldots i_p},
\label{form:tensor representation}
\end{equation}
where $g=(g_{ij})\in \mbox{SO}(3)$ is a rotation in $\mathbb R^3$, defines a linear transformation in a vector space of covariant tensors of rank $p$, i.e. it defines a representation of the rotation group, which is called a tensor representation. A linear transformation (\ref{form:tensor representation}) will be denoted by $g\cdot T$, that is, $\Tilde{T}=g\cdot T$. In this section we give an explicit description of an irreducible 5-dimensional tensor representation of the rotation group in the complex vector space of covariant tensors of rank 3.

Let $\mathcal T^3$ be the vector space of tensors of rank 3 which satisfy the following conditions:
\begin{itemize}
    \item[T1.] A contraction of a tensor $T=(T_{ijk})$ over any pair of subscripts (trace) is zero, that is, for any $j=1,2,3$ it holds
    $$
    T_{iij}=0,\;\;T_{iji}=0,\;\;T_{jii}=0;
    $$
    \item[T2.] For any combination of integers $i,j,k$ (each running from 1 to 3) the sum of the components of tensor $T=(T_{ijk})$, obtained by cyclic permutation of its subscripts, is equal to zero, that is,
    \begin{equation}
    T_{ijk}+T_{jki}+T_{kij}=0.\label{form:sum of cyclic permutations}
    \end{equation}
\end{itemize}
It can be easily verified that the conditions $\mbox{T}_1,\mbox{T}_2$ are invariant under the action of the rotation group (\ref{form:tensor representation}). Hence for any rotation $g\in\mbox{SO}(3)$ we have
$R_g:{\mathcal T}^3\to {\mathcal T}^3$. It is shown in \cite{Gelfand-Minlos-Shapiro} that the vector space ${\mathcal T}^3({\mathbb R}^3)$ is 10-dimensional and the formula (\ref{form:tensor representation}) defines a two-fold irreducible tensor representation of the rotation group in this vector space. If we split the 10-dimensional vector space ${\mathcal T}^3$ into a direct sum of two 5-dimensional subspaces in a way invariant with respect to the action of the rotation group (\ref{form:tensor representation}) then in each 5-dimensional subspace of ${\mathcal T}^3$ we will have an irreducible tensor representation of the rotation group.

One can split the 10-dimensional vector space $\spaceoftensors$ into a direct sum of two 5-dimensional subspaces, which are invariant with respect to a tensor representation of the rotation group, by making use of a linear operator induced by a substitution. Let us denote by $\sigma$ the cyclic substitution of first three integers $\sigma(1)=2, \sigma(2)=3, \sigma(3)=1$. Then one can define the operator $\Phi_\sigma:T\to\Tilde{T}$ acting on the tensors of rank 3 as follows
$$
\Tilde{T}_{i_1i_2i_3}=\Phi_\sigma(T_{i_1i_2i_3})=T_{i_{\sigma(1)} i_{\sigma(2)} i_{\sigma(3)}}=
           T_{i_2i_3i_1},
$$
and extend it by linearity to the vector space of all tensors of rank three. Obviously $\Phi_\sigma^3={\bf 1}$ and
\begin{equation}
({\bf 1}+\Phi_\sigma+\Phi^2_\sigma)\,(T_{ijk})=T_{ijk}+T_{jki}+T_{kij},
\end{equation}
where $\bf 1$ is the identity mapping. Thus the equation (\ref{form:sum of cyclic permutations}) can be written in the form
\begin{equation}
({\bf 1}+\Phi_\sigma+\Phi^2_\sigma)(T)=0.
\label{form:equation equivalent T2}
\end{equation}
Now it is easy to show that the vector space $\spaceoftensors$ is invariant under the action of the operator $\Phi_\sigma$, that is, $\Phi_\sigma:\spaceoftensors\to\spaceoftensors$. Assume that a tensor $T=(T_{ijk})$ satisfies the condition $\mbox{T}_2$ or, equivalently, the equation (\ref{form:equation equivalent T2}). Denote $\Tilde{T}=\Phi_\sigma(T)$. Then
$$
({\bf 1}+\Phi_\sigma+\Phi^2_\sigma)\,(\Tilde{T})=
     ({\bf 1}+\Phi_\sigma+\Phi^2_\sigma)\,(\Phi_\sigma({T}))=
        ({\bf 1}+\Phi_\sigma+\Phi^2_\sigma)\,({T})=0,
$$
and $\Tilde{T}$ also satisfies the equation (\ref{form:equation equivalent T2}). Similarly one can verify that the operator $\Phi_\sigma$ preserves the condition $\mbox{T}_1$.

Generally the property of the linear operator $\Phi_\sigma^3={\bf 1}$ implies that it has three eigenvalues $1,q,\bar q$ in the vector space of all tensors of rank 3. Here $q=\exp{(2i\pi/3)}$ is the primitive third order root of unity and $\bar q$ is its complex conjugate. Another general formula is based on the property of the third order roots of unity $1+q+\bar q=0$. Indeed it is easy to see that due to the mentioned property of the third order roots of unity any tensor of rank 3 can be decomposed into the sum of three tensors
\begin{equation}
T=T_1+T_q+T_{\bar q},
\label{form:decomposition}
\end{equation}
where
\begin{eqnarray}
T_1 &=& \frac{1}{3}({\bf 1}+\Phi_\sigma+\Phi_\sigma^2)(T)\;\;\;\;\;\,\;\;\mbox{or}\;\;\;
            (T_1)_{ijk}=\frac{1}{3}(T_{ijk}+T_{jki}+T_{kij}),\nonumber\\
T_q &=& \frac{1}{3}({\bf 1}+{\bar q}\,\Phi_\sigma+q\,\Phi_\sigma^2)(T)\;\;\;\mbox{or}\;\;\;
            (T_q)_{ijk}=\frac{1}{3}(T_{ijk}+{\bar q}\,T_{jki}+q\,T_{kij}),\nonumber\\
T_{\bar q} &=& \frac{1}{3}({\bf 1}+q\,\Phi_\sigma+{\bar q}\,\Phi_\sigma^2)(T)\;\;\;\mbox{or}\;\;\;
             (T_{\bar q})_{ijk}=\frac{1}{3}(T_{ijk}+{q}\,T_{jki}+{\bar q}\,T_{kij}).\nonumber
\end{eqnarray}
Obviously the tensors $T_1,T_q,T_{\bar q}$ are the eigenvectors of the linear operator $\Phi_\sigma$ corresponding to the eigenvalues $1,q,\bar q$ respectively. Thus we have
\begin{equation*}
\Phi_\sigma(T_1)=T_1,\;\;\Phi_\sigma(T_q)=q\;T_q,\;\;\Phi_\sigma(T_{\bar q})={\bar q}\;T_{\bar q},
\end{equation*}
or, equivalently,
\begin{equation*}
(T_1)_{ijk}=(T_1)_{jki},\;\;(T_q)_{ijk}=\bar q\;(T_q)_{jki},\;\;
                (T_{\bar q})_{ijk}=q\;(T_{\bar q})_{jki}.
\end{equation*}
It is worth to mention that the components $T_q$ and $T_{\bar q}$ of a tensors $T$ satisfy the condition $\mbox{T}_2$.
Restricting (\ref{form:decomposition}) to the vector space $\spaceoftensors$, we see that due to the condition $\mbox{T}_2$ the first term at the right-hand side vanishes, i.e. $T_1=0$ and (\ref{form:decomposition}) takes on the form $T=T_q+T_{\bar q}$, where $T_q,T_{\bar q}\in\spaceoftensors$. Hence we can decompose the vector space $\spaceoftensors$ into the direct sum of two subspaces, which will be denoted by $\spaceoftensorsq$ and $\spaceoftensorsbarq$. Here $\spaceoftensorsq$ is the subspace of the eigenvectors of the linear operator $\Phi_\sigma$ with eigenvalue $q$ and $\spaceoftensorsbarq$ is the subspace of the eigenvectors of the linear operator $\Phi_\sigma$ with eigenvalue $\bar q$. Thus $\spaceoftensors=\spaceoftensorsq\oplus\spaceoftensorsbarq$.

The subspaces $\spaceoftensorsq,\spaceoftensorsbarq$ play a basic role in what follows and it is useful to give here their exact description. $\spaceoftensorsq$ is a vector space of complex-valued tensors of rank 3 which satisfy the condition $\mbox{T}_1$ (trace over any pair of subscripts is zero) and they are eigenvectors of the linear operator $\Phi_\sigma$ with eigenvalue $q$, that is, they satisfy $\Phi_\sigma(T)=q\;T$ or $T_{ijk}={\bar q}\;T_{jki}$. Similarly  $\spaceoftensorsbarq$ is a vector space of complex-valued tensors of rank 3 which satisfy $\mbox{T}_1$ and they are the eigenvectors of the linear operator $\Phi_\sigma$ with eigenvalue $\bar q$, i.e. $\Phi_\sigma(T)={\bar q}\;T$ or $T_{ijk}=q\;T_{jki}$. Hence
\begin{equation}
\spaceoftensorsq=\{T\in\spaceoftensors: \Phi_\sigma(T)=q\;T\},\;\;
   \spaceoftensorsbarq=\{T\in\spaceoftensors: \Phi_\sigma(T)={\bar q}\;T\}.
\end{equation}
The important role of these subspaces is that they are spaces of a 5-dimensional irreducible representation of the rotation group.

A tensor of the third rank $T=(T_{ijk})$ is a quantity with three subscripts $i,j,k$. Therefore, in what follows, it will be convenient for us to represent tensors of the third rank in the form of 3-dimensional matrices, which are also called hypermatrices. By a 3-dimensional matrix, we mean a 3-dimensional cube with components of a tensor $T_{ijk}$ located on the sections of this cube. Here by section we mean a section of a cube by plane perpendicular to its edges. We assume that a cube is located in space so that the first subscript $i$ of a tensor $T_{ijk}$ enumerates sections of a cube parallel to the plane of this page and the numbering starts from the section closest to us ($i=1$) and then takes values 2,3 as the distance from us increases (see figure).
\begin{figure}[htp]
\centering
  \includegraphics[width=9cm]{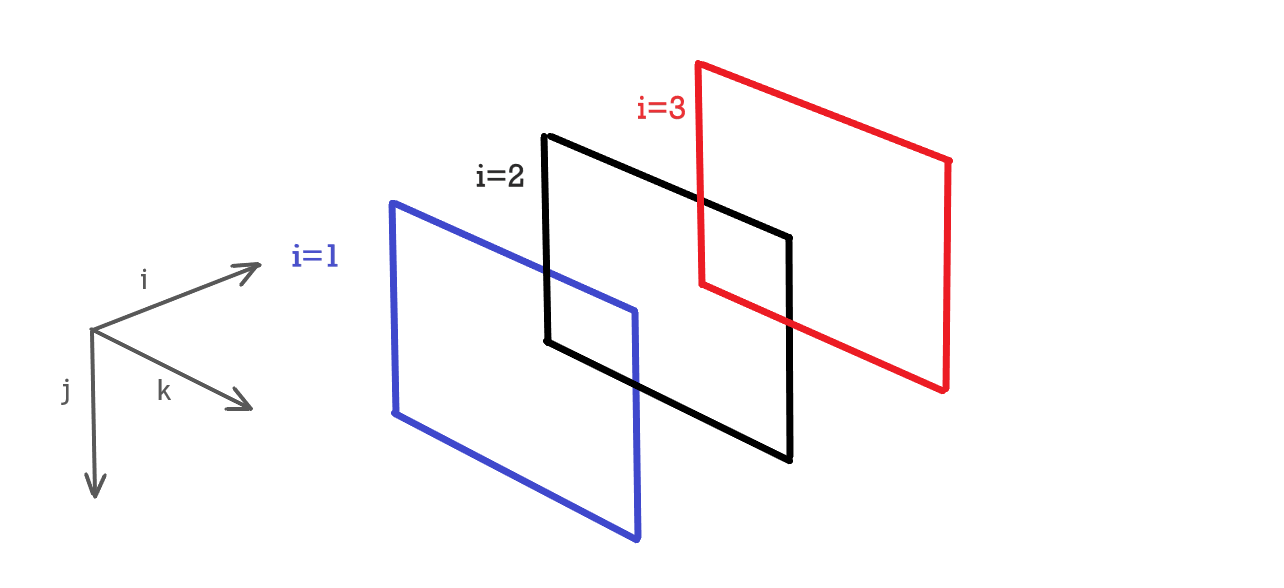}\\
\end{figure}
We will map a 3-dimensional matrix onto the plane of page of this paper by placing the numbered sections of a cube (which are the usual third-order matrices) from left to right, that is, on the left there will be the section with $i=1$, in the center with $i=2$ and on the right with $i=3$. Thus a 3-dimensional matrix of a third order tensor $T=(T_{ijk})$ can be represented as follows
\begin{equation}
T=\left(
  \begin{array}{ccc}
    T_{\textcolor{red}{1}11} & T_{\textcolor{red}{1}12} & T_{\textcolor{red}{1}13} \\
    T_{\textcolor{red}{1}21} & T_{\textcolor{red}{1}22} & T_{\textcolor{red}{1}23} \\
    T_{\textcolor{red}{1}31} & T_{\textcolor{red}{1}32} & T_{\textcolor{red}{1}33} \\
  \end{array}
\left|
\begin{array}{ccc}
    T_{\textcolor{red}{2}11} & T_{\textcolor{red}{2}12} & T_{\textcolor{red}{2}13} \\
    T_{\textcolor{red}{2}21} & T_{\textcolor{red}{2}22} & T_{\textcolor{red}{2}23} \\
    T_{\textcolor{red}{2}31} & T_{\textcolor{red}{2}32} & T_{\textcolor{red}{2}33} \\
  \end{array}
\right|
  \begin{array}{ccc}
    T_{\textcolor{red}{3}11} & T_{\textcolor{red}{3}12} & T_{\textcolor{red}{3}13} \\
    T_{\textcolor{red}{3}21} & T_{\textcolor{red}{3}22} & T_{\textcolor{red}{3}23} \\
    T_{\textcolor{red}{3}31} & T_{\textcolor{red}{3}32} & T_{\textcolor{red}{3}33} \\
  \end{array}
\right),
\label{form:i-direction}
\end{equation}
where a number of a section $i$ is indicated by red color. If a 3-dimensional matrix $T$ is represented in the form (\ref{form:i-direction}) then we will say that $T=(T_{ijk})$ is written in the direction $i$. Analogously we can define $j$-directional and $k$-directional representations of a 3-dimensional matrix.

Now we consider the 5-dimensional complex vector space $\mathbb C^5$ endowed with the canonical Hermitian metric $h$. The coordinates of this space will be denoted by $z^A$, where $A$ runs from 1 to 5. Then $h(z,\bar z)=\sum_{A=1}^5 z^A\,{\bar z}^A$. We will identify this 5-dimensional complex vector space with the complex vector space of the third-rank covariant tensors (or 3-dimensional matrices) $\mathfrak T^3_{\bar q}$ by identifying a point $(z^A)\in {\mathbb C}^5$ with the 3-dimensional matrix $T(z)$, i.e.
$$
(z^A)\in{\mathbb C}^5\;\to\; T(z)=\big(T_{ijk}(z)\big)\in{\mathfrak T}^3_{\bar q},
$$
where
\begin{equation}
T(z)=\left(
  \begin{array}{ccc}
    0 & -\frac{{\bar q}\,z^2}{\sqrt{6}} & \frac{{\bar q}\,z^3}{\sqrt{6}} \\[0.2cm]
    -\frac{q\,z^2}{\sqrt{6}} & \frac{z^1}{\sqrt{6}} & \frac{\,z^4}{\sqrt{3}}\\[0.2cm]
    \frac{q\,z^3}{\sqrt{6}} & \frac{q\,z^5}{\sqrt{3}} & -\frac{z^1}{\sqrt{6}} \\
  \end{array}
\left|
\begin{array}{ccc}
 -\frac{z^2}{\sqrt{6}} & \frac{q\,z^1}{\sqrt{6}} & \frac{{\bar q}\,z^5}{\sqrt{3}} \\[0.2cm]
    \frac{{\bar q}\,z^1}{\sqrt{6}} & 0 & -\frac{{\bar q}\,z^3}{\sqrt{6}}\\[0.2cm]
    \frac{{\bar q}\,z^4}{\sqrt{3}} & -\frac{q\,z^3}{\sqrt{6}} & \frac{z^2}{\sqrt{6}} \\
  \end{array}
\right|
  \begin{array}{ccc}
  \frac{z^3}{\sqrt{6}} & \frac{q\,z^4}{\sqrt{3}} & -\frac{q\,z^1}{\sqrt{6}} \\[0.2cm]
    \frac{z^5}{\sqrt{3}} & -\frac{z^3}{\sqrt{6}} & \frac{q\,z^2}{\sqrt{6}}\\[0.2cm]
    -\frac{{\bar q}\,z^1}{\sqrt{6}} & \frac{{\bar q}\,z^2}{\sqrt{6}} & 0 \\
  \end{array}
\right).
\label{form:T(z)}
\end{equation}
It is easy to verify that $T(z)$ satisfies the condition $\mbox{T}_1$, that is, the trace over any pair of subscripts is zero, and $T_{ijk}(z)=q\,T_{jki}(z).$ Thus, in what follows we will consider the 5-dimensional complex vector space whose points are identified with 3-dimensional complex matrices $T(z)$.

The formula (\ref{form:tensor representation}) defines an action $T\to g\cdot T$ of the rotation group $\mbox{SO}(3)$ on our 5-dimensional complex vector space and this action yields an irreducible tensor representation of the rotation group \cite{Gelfand-Minlos-Shapiro}. Now our aim is to find and study the invariants of this representation, which we will use to construct an irreducible special geometry.

In this paper, we use the classification of invariants of third-order tensors that transform according to formula (\ref{form:tensor representation}), that is, under the action of the rotation group $\mbox{SO}(3)$. If we do not assume that a tensor $T=(T_{ijk})$ has any symmetries, then there is only one linear invariant
\begin{equation*}
I=\epsilon_{ijk}\,T_{ijk}=T_{123}+T_{231}+T_{312}-T_{321}-T_{213}-T_{132},
\end{equation*}
where $\epsilon_{ijk}$ is the Levi-Civita tensor in 3-dimensional Euclidean space. Since we consider the space of tensors that satisfy the property $\mbox{T}_2$, that is, the sum of the components obtained by cyclic permutations of subscripts is equal to zero, the linear invariant $I_1$ vanishes.

The complete set of quadratic $\mbox{SO}(3)$-invariants of a third-order real-valued tensor $T$ (no symmetries) includes eleven invariants. Six of eleven invariants contain the trace of a tensor $T$ with respect to some pair of subscripts, and, due to property $\mbox{T}_1$, these invariants vanish when restricted to the space ${\mathfrak T}^3_{\bar q}$. This leaves us with five $\mbox{SO}(3)$-invariants of a real-valued tensor, and these invariants are on the left side of the table shown below. Due to the fact that we are considering complex-valued tensors, this list of five invariants should be extended by supplementing it with additional invariants. These additional invariants are constructed from those in the left side of the table by replacing one of the factors in a product of tensor components with the complex conjugate and the additional invariants are shown in the right side of the table. Direct calculation leads to the following table of invariants
\begin{eqnarray}
&&\!\!\!\!\!\! I_1=T_{ijk}\,T_{ijk}=0,\qquad\qquad I^{\ast}_1=T_{ijk}\,{\overline{T}}_{ijk}=\sum_{A=1}^5\;z^A\,{\bar z}^A=h(z,\bar z),\\
&&\!\!\!\!\!\! I_2=T_{ijk}\,T_{ikj}=\sum_{A=1}^3\,(z^A)^2+2\,q\,z^4z^5,\;
        I^{\ast}_2=T_{ijk}\,\overline{T}_{ikj}=0,\label{form:I_2}\\
&&\!\!\!\!\!\! I_3=T_{ijk}\,T_{jik}=q\,T_{ijk}\,T_{ikj}=q\;I_2,
   I^\ast_3=T_{ijk}\,\overline{T}_{jik}={\bar q}\,T_{ijk}\,\overline{T}_{ikj}={\bar q}\,I^{\ast}_2=0,\nonumber\\
&&\!\!\!\!\!\! I_4=T_{ijk}\,T_{kji}={\bar q}\,T_{ijk}T_{ikj}={\bar q}\;I_2, I^\ast_4=T_{ijk}\overline{T}_{kji}=q\;T_{ijk}\overline{T}_{ikj}=q\;I^\ast_2=0,\nonumber\\
&&\!\!\!\!\!\! I_5=T_{ijk}T_{kij}+T_{ijk}T_{jki}=0,
       I_5^\ast=T_{ijk}\overline{T}_{kij}+T_{ijk}\overline{T}_{jki}=-I^\ast_1.\nonumber
\end{eqnarray}
The table of invariants shows that we have two independent quadratic invariants $I^\ast_1,I_2$ , where the first one $I^\ast_1$ is the canonical Hermitian metric of the complex five-dimensional space $\mathbb C^5$. Hence each rotation of the 3-dimensional space $\mathbb R^3$ induces a unitary transformation of the complex five-dimensional Hermitian space $\mathbb C^5$, that is, we have a representation ${R}:g\in\mbox{SO}(3)\to R_g\in\mbox{U}(5)$. Obviously the homomorphism $R$ from the rotation group into the group of unitary matrices of order 5 is injective. At the infinitesimal level, the representation $R$ generates the representation of the Lie algebra of the rotation group $\varrho:\tau\in\mathfrak{so}(3)\to\varrho_\tau\in\mathfrak{u}(5)$. Our next goal is to find an explicit form of this representation using the basis of the 5-dimensional complex Hermitian space of 3-dimensional matrices (\ref{form:T(z)}). By other words, given a skew-symmetric third-order matrix $\tau=(\tau_{ij})\in\mathfrak{so}(3)$ we will calculate a fifth-order skew-Hermitian traceless matrix $\varrho_\tau=\big((\varrho_\tau)^A_B\big)\in\mathfrak{u}(5)$. We will see that the form of a skew-Hermitian matrix $\varrho_\tau$ is determined by the second invariant $I_2$, and this matrix is surprisingly similar to the matrix used in the Georgie-Glashow model for unification of elementary particles.

In order to calculate the infinitesimal part of the representation ${R}:\mbox{SO}(3)\to\mbox{U}(5)$, we use the exponential map from the Lie algebra $\mathfrak{so}(3)$ to the rotation group $\mbox{SO}(3)$, taking only the linear part of the corresponding expansion $g_{ij}=\delta_{ij}+\tau_{ij}+\ldots$, where $\tau=(\tau_{ij})$ is a skew-symmetric matrix. We can write
\begin{eqnarray}
(g\cdot T)_{ijk} &=& g_{ip}g_{jr}g_{ks} T_{prs}
   =(\delta_{ip}+\tau_{ip}+\ldots) (\delta_{jr}+\tau_{jr}+\ldots)(\delta_{ks}+\tau_{ks}+\ldots)T_{prs}\nonumber\\
     &=& T_{ijk}+\tau_{ip}T_{pjk}+\tau_{jr}T_{irk}+\tau_{ks}T_{ijs}+\ldots
\end{eqnarray}
Hence the infinitesimal part of the action $T\to g\cdot T$ (up to the terms of the second order and higher) defines the linear operator $\varrho_\tau:{\mathbb C}^5\to{\mathbb C}^5$, where
\begin{equation}
(\varrho_\tau(T))_{ijk}=\tau_{ip}T_{pjk}+\tau_{jr}T_{irk}+\tau_{ks}T_{ijs}.
\label{form:infinitesimal action}
\end{equation}
It will be convenient for us to pass to a parameterization of matrix $\tau=(\tau_{ij})$ with the help of parameters containing one index. Let us define $\tau_i=-\frac{1}{2}\epsilon_{ijk}\tau_{jk}$.  Then
$$
\tau=\left(\begin{array}{ccc}
    0 & -\tau_3 & \tau_2 \\
    \tau_3 & 0 & -\tau_1 \\
    -\tau_2 & \tau_1 & 0 \\
\end{array}
\right)
$$
Now we can calculate a matrix of this operator (we will use the same notation $\varrho_\tau$ for the matrix of the operator) by means of the following basis in 5-dimensional complex space of 3-dimensional matrices (\ref{form:T(z)})
\begin{eqnarray}
&&\left(
  \begin{array}{ccc}
    0\! &\! 0 \!&\!\! 0 \!\\[0.2cm]
    0\! &\! 1 \!&\!\! 0 \!\\[0.2cm]
    0\! &\! 0 \!&\!\! -1 \!\\
  \end{array}
\left|
\begin{array}{ccc}
 0\! &\! q \!&\! 0\! \\[0.2cm]
 {\bar q} \!&\! 0 \!&\! 0\!\\[0.2cm]
    0 \!&\! 0 \!&\! 0 \!\\
  \end{array}
\right|
  \begin{array}{ccc}
  \!0 \!&\! 0 \!&\!\! -q \!\\[0.2cm]
  \!0 \!&\! 0 \!&\!\! 0 \!\\[0.2cm]
 \!-{\bar q} \!&\!\! 0 \!&\! 0 \!\\
  \end{array}
\right),
\;\;\left(
  \begin{array}{ccc}
   \! 0 \!&\!\! -{\bar q} \!&\! 0 \!\\[0.2cm]
   \! -q \!&\!\! 0 \!&\! 0\\[0.2cm]
   \! 0 \!&\!\! 0 \!&\! 0 \!\\
  \end{array}
\left|
\begin{array}{ccc}
 \!-1 \!&\! 0 \!&\! 0 \!\\[0.2cm]
\! 0 \!&\! 0 \!&\! 0 \!\\[0.2cm]
  \!  0 \!&\! 0 \!&\! 1 \!\\
  \end{array}
\right|
  \begin{array}{ccc}
  0 \!&\! 0 \!&\! 0 \!\\[0.2cm]
  0 \!&\! 0 \!&\! q\\[0.2cm]
 0 \!&\! {\bar q} \!&\! 0 \!\\
  \end{array}
\right),\nonumber\\
&&\left(
  \begin{array}{ccc}
    0\! &\! 0 \!&\!\! {\bar q} \!\\[0.2cm]
    0\! &\! 0 \!&\!\! 0 \!\\[0.2cm]
    q\! &\! 0 \!&\!\! 0 \!\\
  \end{array}
\left|
\begin{array}{ccc}
 0\! &\! 0 \!&\! 0\! \\[0.2cm]
 0 \!&\! 0 \!&\! -{\bar q}\!\\[0.2cm]
    0 \!&\! -q \!&\! 0 \!\\
  \end{array}
\right|
  \begin{array}{ccc}
  \!1 \!&\! 0 \!&\!\! 0 \!\\[0.2cm]
  \!0 \!&\! -1 \!&\!\! 0 \!\\[0.2cm]
 \! 0 \!&\!\! 0 \!&\! 0 \!\\
  \end{array}
\right),
\left(
  \begin{array}{ccc}
   \! 0 \!&\!\! 0 \!&\! 0 \!\\[0.2cm]
   \! 0 \!&\!\! 0 \!&\! \sqrt{2}\\[0.2cm]
   \! 0 \!&\!\! 0 \!&\! 0 \!\\
  \end{array}
\left|
\begin{array}{ccc}
 \!0 \!&\! 0 \!&\! 0 \!\\[0.2cm]
\! 0 \!&\! 0 \!&\! 0 \!\\[0.2cm]
  \!  \sqrt{2}{\bar q} \!&\! 0 \!&\! 0 \!\\
  \end{array}
\right|
  \begin{array}{ccc}
  0 \!&\! \sqrt{2}q \!&\! 0 \!\\[0.2cm]
  0 \!&\! 0 \!&\! 0\\[0.2cm]
 0 \!&\! 0 \!&\! 0 \!\\
  \end{array}
\right),\nonumber\\
&&
\qquad\qquad\qquad
\left(
  \begin{array}{ccc}
   \! 0 \!&\!\! 0 \!&\! 0 \!\\[0.2cm]
   \! 0 \!&\!\! 0 \!&\! 0\\[0.2cm]
   \! 0 \!&\!\! \sqrt{2}q \!&\! 0 \!\\
  \end{array}
\left|
\begin{array}{ccc}
 \!0 \!&\! 0 \!&\! \sqrt{2}{\bar q} \!\\[0.2cm]
\! 0 \!&\! 0 \!&\! 0 \!\\[0.2cm]
  \!  0 \!&\! 0 \!&\! 0 \!\\
  \end{array}
\right|
  \begin{array}{ccc}
  0 \!&\! 0 \!&\! 0 \!\\[0.2cm]
  \sqrt{2} \!&\! 0 \!&\! 0\\[0.2cm]
 0 \!&\! 0 \!&\! 0 \!\\
  \end{array}
\right),\nonumber
\end{eqnarray}
Let us enumerate the 3-dimensional matrices of this basis (starting from the left in the first row and moving from left to right and then from top to bottom) as follows ${E}_A$, where $A=1,2,\ldots, 5$. By straightforward calculation we find
\begin{eqnarray}
\varrho_\tau(E_1) &=& -\tau_3\,{E}_2 + \tau_2\,{E}_3 + \sqrt{2}\tau_1\,{E}_4 + \sqrt{2}\,{\bar q}\,\tau_1\,E_5,\nonumber\\
\varrho_\tau(E_2) &=& \tau_3\,{E}_1 - \tau_1\,{E}_3 + \sqrt{2}\,q\,\tau_2\,{E}_4 + \sqrt{2}\,{q}\,\tau_2\,{E}_5,\nonumber\\
\varrho_\tau(E_3) &=& -\tau_2\,{E}_1 + \tau_1\,{E}_2 + \sqrt{2}\,{\bar q}\,\tau_3\,{E}_4 + \sqrt{2}\,\tau_3\,{E}_5,\nonumber\\
\varrho_\tau({E}_4) &=& -\sqrt{2}\,\tau_1\,{E}_1 - \sqrt{2}\,{\bar q}\,\tau_2\,{E}_2 - \sqrt{2}\,{q}\,\tau_3\,{E}_3,\nonumber\\
\varrho_\tau({E}_5) &=& -\sqrt{2}\,q\,\tau_1\,{E}_1 - \sqrt{2}\,{\bar q}\,\tau_2\,{E}_2 - \sqrt{2}\,\tau_3\,{E}_3.\nonumber
\end{eqnarray}
Hence the matrix of the operator $\varrho_\tau$ has the form
\begin{equation}
\varrho_\tau=\left(
     \begin{array}{ccccc}
       0   & \tau_3 & -\tau_2 & -\sqrt{2}\,\tau_1 & -\sqrt{2}\,q\,\tau_1  \\
       -\tau_3   & 0 & \tau_1 & -\sqrt{2}\,{\bar q}\,\tau_2 & -\sqrt{2}\,{\bar q}\,\tau_2  \\
       \tau_2   & -\tau_1 & 0 & -\sqrt{2}\,q\,\tau_3 & -\sqrt{2}\,\tau_3  \\[0.2cm]
       \sqrt{2}\,\tau_1   & \sqrt{2}\,q\,\tau_2 & \sqrt{2}\,{\bar q}\,\tau_3 & 0 & 0  \\
       \sqrt{2}\,{\bar q}\,\tau_1  & \sqrt{2}\,{q}\,\tau_2 & \sqrt{2}\,\tau_3 & 0 & 0  \\
     \end{array}
     \right).
    \label{form:infinitesimal matrix}
\end{equation}
Due to the fact that the irreducible representation of the rotation group in the complex space of three-dimensional matrices (\ref{form:T(z)}) is unitary (as we mentioned above one of the invariants of this representation is the Hermitian metric of the 5-dimensional complex space), the matrix of the representation of the Lie algebra of the rotation group $\varrho_\tau$ must be skew-Hermitian, and this is indeed the case, because the matrix $\varrho_{\tau}$ satisfies the relation $\varrho_\tau+\varrho_\tau^\dag=0$, where $\varrho_\tau^\dag={\overline \varrho}_\tau^{\tt t}$. It is easy to see that $\rho_\tau$ is a traceless matrix. Hence $\rho_\tau$ belongs to the Lie algebra of the group $\mbox{SU}(5)$, that is, $\rho_\tau\in\frak{su}(5)$. Hence we can express this matrix in terms of generators of $\frak{su}(5)$ denoted in physics papers by $L_i$, where $i=1,2,\ldots,24$, and $L_i$ are Hermitian traceless matrices of 5th order normalized by $\mbox{Tr}(L_i\,L_j)=\frac{1}{2}\,\delta_{ij}$. In this paper we use the following numbering of the generators of $\frak{su}(5)$:
\begin{itemize}
\item
The first eight generators correspond to $\mbox{SU}(3)$, that is,
$$
L_k=\frac{1}{2}\left(\begin{array}{cc}
\lambda_k & 0\\
0         & 0\\
\end{array}\right),
$$
where $\lambda_i$ are Gell-Mann matrices,
\item
the next four generators $L_{9},L_{10}, L_{11}, L_{12}$ have the form
$$
L_{8+k}=\frac{1}{2}\left(\begin{array}{cc}
0 & 0\\
0 & \sigma_k\\
\end{array}\right),\;\; L_{12}=\frac{1}{2\sqrt{15}}\,\mbox{Diag}(-2,-2,-2,3,3),
$$
where $k=1,2,3$ and $\sigma_1,\sigma_2,\sigma_3$ are Pauli matrices,
\item
the next twelve generators (sometimes called broken matrices) are of the form
\begin{eqnarray}
L_{12+k} &=& \frac{1}{2}(d^k_4+d^4_k),\;\;\;\;\;\;L_{15+k}=\frac{1}{2}(d^k_5+d^5_k),\nonumber\\
    L_{18+k} &=& -\frac{i}{2}(d^k_4-d^4_k),\;\;\;L_{21+k}=-\frac{i}{2}(d^k_5-d^5_k),\nonumber
\end{eqnarray}
where $k=1,2,3$ and $d^i_k$ is a matrix with only one non-zero element, which is at the intersection of $i$th row with $k$th column.
\end{itemize}
Then the matrix $\rho_\tau$ can be written in the terms of $\frak{su}(5)$-generators $L_i$ as follows
\begin{eqnarray}
\rho_\tau \!\!\!&=&\!\!\! 2i\,\tau_1\,(L_7-\frac{\sqrt{6}}{2}L_{16}-\sqrt{2}\,L_{19}+\frac{\sqrt{2}}{2}L_{22})\nonumber\\
     &&\;\;\;+\;
    2i\,\tau_2\,(-L_5+\frac{\sqrt{6}}{2}L_{14}-\frac{\sqrt{6}}{2}\,L_{17}+\frac{\sqrt{2}}{2}L_{20}+\frac{\sqrt{2}}{2}L_{23})\nonumber\\
      &&\;\;\;\;\;\;\;\;+2i\,\tau_3\,(L_2-\frac{\sqrt{6}}{2}\,L_{15}+\frac{\sqrt{2}}{2}\,L_{21}-\sqrt{2}\,L_{24}).\nonumber
\end{eqnarray}
It should be noted here that the matrix $\rho_\tau$ is not only skew-Hermitian and traceless, it also satisfies some additional conditions that follow from the fact that the irreducible representation of the rotation group has one more quadratic invariant $I_2$ (\ref{form:I_2}). We will denote the quadratic form in the 5-dimensional complex vector space induced by this invariant as follows
\begin{equation}
{K(z,z)}={K}_{AB}\;z^A\,z^B=(z^1)^2 + (z^2)^2 + (z^3)^2   ` + 2\,q\,z^4\,z^5.
\end{equation}
The matrix of this quadratic form
\begin{equation}
K=({K}_{AB})=\left(
                     \begin{array}{ccccc}
                1 & 0 & 0 & 0 & 0 \\
                0 & 1 & 0 & 0 & 0 \\
                0 & 0 & 1 & 0 & 0 \\
                0 & 0 & 0 & 0 & q \\
                0 & 0 & 0 & q & 0 \\
                     \end{array}\right),
\end{equation}
can be considered as a covariant second-order tensor in the 5-dimensional complex vector space and the properties of this tensor will be studied in the next section. Here we only note that the matrix $K={(K_{AB})}$ is symmetric and unitary, i.e.
\begin{equation}
{K}={K}^{\tt T},\;\;{K}\,{K}^\dag=E,
\end{equation}
where $E=(\delta_{AB})$ is the identity matrix.

The infinitesimal action (\ref{form:infinitesimal action}) generates the following vector fields in 5-dimensional complex space
\begin{eqnarray}
{\mathtt X}_1 \!\!\!&=&\!\!\! (\sqrt{2}\,z^4 + \sqrt{2}\,q\,z^5)\,\frac{\partial }{\partial z^1}-z^3\,\frac{\partial}{\partial z^2}+z^2\,\frac{\partial}{\partial z^3}-
       \sqrt{2}\,z^1\frac{\partial}{\partial z^4}-\sqrt{2}\,{\bar q}\,z^1\frac{\partial}{\partial z^5},\nonumber\\
{\mathtt X}_2 \!\!\!&=&\!\!\! z^3\frac{\partial }{\partial z^1}+\sqrt{2}\,{\bar q}\,(z^4+z^5)\,\frac{\partial}{\partial z^2}-z^1\,\frac{\partial}{\partial z^3}
       -\sqrt{2}\,q\,z^2\frac{\partial}{\partial z^4}-\sqrt{2}\,{q}\,z^2\frac{\partial}{\partial z^5},\nonumber\\
{\mathtt X}_2 \!\!\!&=&\!\!\! -z^2\frac{\partial }{\partial z^1}+z^1\,\frac{\partial}{\partial z^2}+\sqrt{2}(q\,z^4+z^5)\,\frac{\partial}{\partial z^3}
       -\sqrt{2}\,{\bar q}\,z^3\frac{\partial}{\partial z^4}-\sqrt{2}\,z^3\,\frac{\partial}{\partial z^5}.\nonumber
\end{eqnarray}
These vector fields span the Lie algebra $[{\mathtt X}_1,{\mathtt X}_2]={\mathtt X}_3,[{\mathtt X}_2,{\mathtt X}_3]={\mathtt X}_1,[{\mathtt X}_3,{\mathtt X}_1]={\mathtt X}_2$ isomorphic to the Lie algebra of matrices (\ref{form:infinitesimal matrix}). Due to the fact that the Hermitian metric $h(z,{\bar z})$ and the quadratic form $K(z)$ are invariants of the irreducible representation of the rotation group $g\in\mbox{SO}(3)\to R_g\in\mbox{U}(5)$, the vector fields ${\mathtt X}_1,{\mathtt X}_2,{\mathtt X}_3$ vanish on the Hermitian form $h(z,\bar z)$ and the quadratic form $K(z)$.

Now our goal is to show that, in fact, the irreducible representation of the rotation group $g\in\mbox{SO}(3)\to R_g\in\mbox{U}(5)$ has the form $g\in\mbox{SO}(3)\to R_g\in\mbox{SU}(5)$, that is, each rotation generates a special (with determinant equal to 1) unitary transformation in the 5-dimensional complex vector space. For this purpose, we will find a parameterization of the irreducible representation using Euler angles. Let us consider two one-parameter subgroups of the rotation group
\begin{equation}
g_1(t)=\left(
                      \begin{array}{ccc}
                        \cos t & -\sin t & 0 \\
                        \sin t & \cos t & 0 \\
                        0 & 0 & 1 \\
                      \end{array}
                    \right),\;\;\;
g_2(t)=\left(
                      \begin{array}{ccc}
                      1 & 0 & 0 \\
                      0 & \cos t & -\sin t \\
                      0 &  \sin t & \cos t \\
                      \end{array}
                    \right).
\end{equation}
The one-parameter subgroups of unitary transformations in 5-dimensional complex vector space generated by the irreducible representation of $g_1(t)$ and $g_2(t)$ have the following form respectively
\begin{eqnarray}
R_1(t) &=& \left(
                      \begin{array}{ccccc}
                        \cos t & \sin t & 0 & 0 & 0 \\
                        -\sin t & \cos t & 0 & 0 & 0 \\
                        0 & 0 & \cos 2t & -\frac{q}{\sqrt{2}}\sin 2t & -\frac{1}{\sqrt{2}}\sin 2t \\
                        0 & 0 & \frac{\bar q}{\sqrt{2}}\sin 2t & \cos^2 t & -{\bar q}\sin^2 t \\
                        0 & 0 & \frac{1}{\sqrt{2}}\sin 2t & -q\,\sin^2 t & \cos^2 t \\
                      \end{array}
                    \right),\label{form:R_1}\\
R_2(t) &=& \left(
                      \begin{array}{ccccc}
                        \cos 2t & 0 & 0 & -\frac{1}{\sqrt{2}}\sin 2t & -\frac{q}{\sqrt{2}}\sin 2t \\
                        0 & \cos t & \sin t & 0 & 0 \\
                        0 & -\sin t & \cos t & 0 & 0 \\
                        \frac{1}{\sqrt{2}}\sin 2t  & 0 & 0 & \cos^2 t & -{q}\sin^2 t \\[0.3cm]
                        \frac{\bar{q}}{\sqrt{2}}\sin 2t & 0 & 0 & -\bar{q}\,\sin^2 t & \cos^2 t \\
                      \end{array}
                    \right).\label{form:R_2}
\end{eqnarray}
Direct calculation shows that the determinants of these matrices are equal to 1. Since any rotation can be written as a composition $g_1(\phi)\,g_2(\theta)\,g_1(\psi)$, where $\phi,\theta,\psi$ are Euler angles, we conclude that each rotation generates a unitary transformation with determinant 1, that is, the irreducible representation has the form of inclusion $R: \mbox{SO}(3)\hookrightarrow \mbox{SU}(5)$ and we will denote the image of the rotation group with respect to this inclusion by $\mathfrak G_3$. Hence ${\mathfrak G}_3\subset \su5$.
\section{$\mbox{SO}(3)$-irreducible geometric structure on a five-dimensional Hermitian manifold}
The purpose of this section is to study the properties of the quadratic form
\begin{equation}
K(z,z)=(z^1)^2+(z^2)^2+(z^3)^2+2\,qz^4\,z^5,
\end{equation}
which is invariant under the irreducible representation $R$ of the rotation group, where $R:\mbox{SO}(3)\hookrightarrow\mbox{SU}(5)$. In the previous section we denoted the image of this inclusion by $\mathfrak G_3$ and, according to the formulas (\ref{form:R_1}), (\ref{form:R_2}), any element of the group $\mathfrak G_3$ can be written as a product $R_1(t)\,R_2(s)\,R_1(v)$, where $t,s,v$ are real parameters. Hence $\mathfrak G_3$ is a stabilizer of the quadratic form $K(z,z)$ in $\mbox{SU}(5)$. Assume that ${z}^A=U^A_B\,\tilde{z}^B$, where $U=(U^A_B)$ is a regular complex $5\times 5$-matrix, is a linear transformation in the five-dimensional complex space $\mathbb C^5$. Then the matrix of the quadratic form $K(z,z)$
\begin{equation}
K=\left(
                     \begin{array}{ccccc}
                1 & 0 & 0 & 0 & 0 \\
                0 & 1 & 0 & 0 & 0 \\
                0 & 0 & 1 & 0 & 0 \\
                0 & 0 & 0 & 0 & q \\
                0 & 0 & 0 & q & 0 \\
                     \end{array}\right),
\label{form:matrix K2}
\end{equation}
transforms under this transformation as follows
\begin{equation}
{\tilde K}_{AB}=U^C_A\,U^D_B\;K_{CD},\;\;\; {K}={\tilde K}_{AB}{\tilde z}^A{\tilde z}^B,
\label{form:transformation rule}
\end{equation}
or in the matrix form
\begin{equation}
\tilde{K}=U^{\tt T}\,K\,U,
\label{form:unitary transformation of K }
\end{equation}
where $K=(K_{AB}),\tilde K=(\tilde K_{AB})$ are matrices of the form $K(z,z)$ in different bases of the Hermitian space $\mathbb C^5$ and $U^{\tt T}$ is the transposed matrix of $U$. The set of all matrices $\tilde K$ obtained with the help of (\ref{form:unitary transformation of K }) will be referred to as an orbit of the quadratic form $K(z,z)$ with an indication of a group of transformations. For example, the set of all matrices $\tilde K$ obtained by means of unitary transformations will be referred to as a $\mbox{U}(5)$-orbit of $K(z,z)$. Obviously, we can consider the matrix $K=(K_{AB})$ as a second-order covariant tensor in a five-dimensional vector space and in this case we will talk about the $\mbox{U}(5)$-orbit of the tensor $K=(K_{AB})$. Our aim in this section is to find properties of the quadratic form $K(z,z)$ (or of the corresponding tensor $K=(K_{AB})$) such that they will uniquely determine the orbit of this quadratic form.

First of all, it is easy to see that the tensor $K$ is symmetric and unitary and these properties are invariant with respect to the group of unitary transformations $\unitary$. Indeed for any $U\in\unitary$ we have
$$
\tilde{K}^{\tt T}=(U^{\tt T}\,K\,U)^{\tt T}=U^{\tt T}\,K^{\tt T}\,U=U^{\tt T}\,K\,U=\tilde{K},
$$
and
$$
\tilde{K}\,\overline{\tilde{K}}=U^{\tt T}\,K\,(U\;\overline{U}^{\tt T})\,\overline{K}\,\overline{U}=
    U^{\tt T}\,(K\,\overline{K})\,\overline{U}=U^{\tt T}\,\overline{U}=E,
$$
where $\overline{U}$ is the complex conjugate matrix of $U$ and $E$ is the unit matrix. Hence the $\unitary$-orbit of the tensor $K=(K_{AB})$ is an orbit of symmetric and unitary tensor.

We recall that determinant of the matrix of a quadratic form is referred to as a discriminant of a quadratic form. It is easy to find that the discriminant of the quadratic form $K(z,z)$ is $\epsilon$, where $\epsilon=\exp(i\,\pi/3)$ is the primitive sixth-order root of unity. But the discriminant of the quadratic form $K(z,z)$ is invariant with respect to the action of the group $\unitary$. Indeed we have
$$
\mbox{det}\,\tilde{K}=\mbox{det}\,(U^{\tt T}\,K\,U)=(\mbox{det}\, U)^2\,\mbox{det}\,K=\epsilon.
$$
Hence the $\unitary$-orbit of the second order covariant tensor $K=(K_{AB})$ is an orbit of the tensor with determinant equal to $\epsilon$.

The $\mbox{U}(5)$-invariant properties of the tensor $K=(K_{AB})$ found above do not yet uniquely determine the $\unitary$-orbit of this tensor in the space of $\unitary$-orbits of all second order covariant tensors. In order to find additional invariant conditions we use the following fact from the matrix calculus.
It is known \cite{Gantmacher} that a symmetric and unitary complex matrix $X$, that is,
$$
X=X^{\tt T}=\overline{X}^{-1},
$$
can be written in the exponential form $X=e^{iY}$, where $Y$ is the real symmetric matrix. The tensor $K=(K_{AB})$ is symmetric, unitary, $\mbox{det}\,K=\epsilon$ and these properties are $\unitary$-invariant. Thus in any orthonormal basis for the five-dimensional complex space $\mathbb C^5$ or, by other words, at any point of $\unitary$-orbit this tensor considered as a matrix can be written in the exponential form ${K}=\exp(i{S})$, where $S$ is a real symmetric matrix.  Particularly in the case of the matrix (\ref{form:matrix K2}) a straightforward computation gives the block form of the 5th order real symmetric matrix $S$
\begin{equation}
S=\left(
                     \begin{array}{cc}
                0\;\; | &\!\!\! 0 \\
                \hline
                0\;\; | &\!\!\! \Sigma\\
                     \end{array}\right),\;\;\;
\Sigma=\left(\begin{array}{cc}
               \frac{\pi}{6} & \frac{\pi}{2} \\
               \frac{\pi}{2} & \frac{\pi}{6} \\
               \end{array}\right).
\end{equation}

In particular case when $U=(U^A_B)$ is a real unitary transformation, that is $U^{\tt T}=\overline{U}^{-1}=U^{-1}$, we can easily find a transformation law of the matrix $S$. Indeed in the case of a real unitary matrix $U$ we have
$$
\tilde{K}=U^{\tt T}\,K\,U=U^{-1}\,K\,U=U^{-1}\,e^{iS}\,U=e^{i\,(U^{-1}\,S\,U)},
$$
and $\tilde{K}=\exp(i\tilde{S})$ implies $\tilde{S}=U^{-1}\,S\,U$. But a real unitary matrix is an orthogonal real matrix and, making use of a transformation $\tilde{S}=U^{-1}\,S\,U$, the real symmetric matrix $S$ can be put into a diagonal form. Straightforward computation gives the diagonal forms of matrices $\tilde S, \tilde K$
\begin{equation}
\tilde S=\left(
                     \begin{array}{cc}
                0\;\; | &\!\!\! 0 \\
                \hline
                0\;\; | &\!\!\! \tilde\Sigma\\
                     \end{array}\right),\;\;
\tilde \Sigma=\left(
                     \begin{array}{cc}
                -\frac{\pi}{3} & 0 \\
                0 & \frac{2\pi}{3} \\
                     \end{array}\right),\;\;
\tilde K=e^{i\tilde S}=\left(
                     \begin{array}{cc}
                E\;\; | &\!\!\! 0 \\
                \hline
                0\;\; | &\!\!\! \Xi\\
                     \end{array}\right),\;\;
\Xi=\left(
                     \begin{array}{cc}
                \epsilon^5 & 0 \\
                0 & \epsilon^2 \\
                     \end{array}\right),
\label{form:diagonal form of K}
\end{equation}
where $\epsilon^5=-q, \epsilon^2=q$ and $E$ is the 3rd order unit matrix. It is easy to verify that the sixth power of the matrix $K$ of the quadratic form $K(z,z)$ is equal to the identity matrix, that is, $K^6=E$, and this relation is invariant under real unitary transformations of the five-dimensional complex space. It is well known that if the $n$th power of a matrix is equal to the identity matrix, then the eigenvalues of such a matrix are the $n$th roots of unity. Thus, the diagonal form $\tilde K$ of the matrix $K$ in (\ref{form:diagonal form of K}) with the sixth order roots of unity on the main diagonal, is a consequence of the fact that $K$ or $\tilde K$ to the sixth power is equal to the identity matrix. We proved the following statement
\begin{prop}
For any orthonormal basis $\{E_A\}$, where $A=1,2,\ldots,5$, for the five-dimensional complex Hermitian space $\mathbb C^5$ the second order covariant tensor
$K_{AB}=K(E_A,E_B)$ determined by the quadratic form  $K(z,z)=(z^1)^2+(z^2)^2+(z^3)^2+2q\,z^4\,z^5$
has the following $\unitary$-invariant properties:
\begin{itemize}
\item $K_{AB}=K_{BA}$ (symmetric),
\item $K_{AB}\,\overline{K}_{CB}=\delta_{AB}$ (unitary),
\item $\mbox{det}\,(K_{AB})=\epsilon$, where $\epsilon=e^{i\pi/3}$ is the sixth order root of unity.
\end{itemize}
It also has the following properties, which are invariant with respect to real unitary transformations:
\begin{itemize}
\item $K^6=E$, where the tensor $K=(K_{AB})$ is considered as a matrix,
\item the eigenvalues of the tensor $K=(K_{AB})$ are $1,1,1,q,-q$, where $q=e^{2\pi i/3}$ is the cubic root of unity.
\end{itemize}
\label{proposition}
\end{prop}
\noindent
This statement provides a basis for studying five-dimensional complex manifolds with a structure determined by the tensor $K=(K_{AB})$. Let $(M,h)$ be a five-dimensional Hermitian manifold, where $h$ is a Hermitian metric. A Hermitian metric $h$ makes it possible to reduce the group of non-degenerate linear transformations of a tangent space $T_xM, x\in M$ of a manifold $M$ to the group of unitary transformations $\unitary$. In other words, we can consider the principal bundle of orthonormal frames over a manifold $M$ with the structure group $\unitary$. Thus, by a tensor field on a manifold $M$ we mean a tensor defined at each point of a manifold $M$ and transformed under the action of the structure group $\unitary$. If $\mbox{O}_{\mathbb R}(5)\subset\unitary$ is the subgroup of real unitary matrices then we can consider the sub-orbit of a $\unitary$-tensor field, that is, the tensor field transforming according to the action of the subgroup $\mbox{O}_{\mathbb R}(5)$ and this sub-orbit will be referred to as a $\mbox{O}_{\mathbb R}(5)$-tensor.
\begin{defn}
An $\mbox{SO}(3)$-irreducible geometric structure on a five-dimensional complex Hermitian manifold $(M,h)$ is a 2nd order covariant symmetric, unitary tensor field $K_{AB}$ whose determinant is equal to the primitive sixth order root of unity $\epsilon=e^{i\pi/3}.$  Moreover, the tensor field $K_{AB}$ considered as a $\mbox{O}_{\mathbb R}(5)$-tensor field has the eigenvalues $1,q,-q$, where the multiplicity of the eigenvalue $1$ is 3, and $q$ is the primitive cubic root of unity $q=e^{2i\pi/3}$.
\label{definition}
\end{defn}
\noindent
From this definition it follows that an $\mbox{SO}(3)$-irreducible geometric structure on a five-dimensional Hermitian manifold $M$ can be considered as a triple $(M,h,K)$, where $h$ is a Hermitian metric of $M$ and $K$ is a 2nd order covariant tensor field defined on $M$ or the corresponding quadratic form. Two triples $(M,h,K)$ and $(\tilde M, \tilde h,\tilde K)$ will be referred to as equivalent $\mbox{SO}(3)$-irreducible geometric structures on Hermitian manifolds $M,\tilde M$ respectively if there exists a diffeomorphism $\psi:M\to \tilde M$ such that
$$
h(v,w)=\tilde h(\psi_\ast(v),\psi_\ast(w)),\;\;\;K(v,w)=\tilde K(\psi_\ast(v),\psi_\ast(w)),
$$
where $v,w$ are tangent vectors to a manifold $M$, $\psi_\ast$ is the differential of a diffeomorphism $\psi$ and $K,\tilde K$ are quadratic forms induced by the tensors $K_{AB},\tilde K_{AB}$ respectively.

Let us study a local structure of a manifold $M$. It follows from Proposition \ref{proposition} and Definition \ref{definition} that locally we can choose a frame $\{{\tt E}_A\}$ of vector fields ${\tt E}_A$ and its dual coframe $\{\theta^A\}$ of complex-valued 1-forms, i.e. $\theta^A({\tt E}_B)=\delta^A_B$, such that
\begin{itemize}
\item  $\{{\tt E}_A\}$ is an orthonormal frame, that is, $h({\tt E}_A,{\tt E}_B)=\delta_{AB}$ and
$$
h=\theta^1\,\overline{\theta}^1+\theta^2\,\overline{\theta}^2+\theta^3\,\overline{\theta}^3+\theta^4\,\overline{\theta}^4+\theta^5\,\overline{\theta}^5,
$$
\item the components of the tensor $K_{AB}$ form the following matrix
$$
K=\left(
                     \begin{array}{ccccc}
                1 & 0 & 0 & 0 & 0 \\
                0 & 1 & 0 & 0 & 0 \\
                0 & 0 & 1 & 0 & 0 \\
                0 & 0 & 0 & 0 & q \\
                0 & 0 & 0 & q & 0 \\
                     \end{array}\right),
$$
and the quadratic form induced by these components is
\begin{equation}
K=(\theta^1)^2+(\theta^2)^2+(\theta^3)^2+2q\,\theta^4\,\theta^5.
\label{form:quadratic form in thetas}
\end{equation}
\end{itemize}
It is clear that the subgroup $\frak G_3\subset\unitary$ (isomorphic to the rotation group) studied at the end of the previous section is the stabilizer of the quadratic form (\ref{form:quadratic form in thetas}). Hence we can reduce the gauge group $\unitary$ to this subgroup and consider a $\frak g_3$-connection 1-form $\omega$ on a manifold $M$, where $\frak g_3$ is the Lie algebra of $\frak G_3$. We can write this $\frak g_3$-valued connection 1-form as follows
\begin{equation}
\omega=\left(
     \begin{array}{ccccc}
       0   & \omega^3 & -\omega^2 & -\sqrt{2}\,\omega^1 & -\sqrt{2}\,q\,\omega^1  \\
       -\omega^3   & 0 & \omega^1 & -\sqrt{2}\,{\bar q}\,\omega^2 & -\sqrt{2}\,{\bar q}\,\omega^2  \\
       \omega^2   & -\omega^1 & 0 & -\sqrt{2}\,q\,\omega^3 & -\sqrt{2}\,\omega^3  \\[0.2cm]
       \sqrt{2}\,\omega^1   & \sqrt{2}\,q\,\omega^2 & \sqrt{2}\,{\bar q}\,\omega^3 & 0 & 0  \\
       \sqrt{2}\,{\bar q}\,\omega^1  & \sqrt{2}\,{q}\,\omega^2 & \sqrt{2}\,\omega^3 & 0 & 0  \\
     \end{array}
     \right),
    \label{form:connection form}
\end{equation}
where $\omega^1,\omega^2,\omega^3$ are real-valued 1-forms. It is easy to see that a connection 1-form $\omega$ is a skew-Hermitian, that is, $\overline{\omega}^{\tt T}=-\omega$. Then the torsion 2-form ${\tt T^A}$ and the curvature 2-form ${\tt R^A_B}$ of a connection $\omega$ can be expressed as follows
\begin{equation}
{\tt T^A}=d\theta^A+\omega^A_B\wedge\theta^B,\;\;{\tt R^A_B}=d\omega^A_B+\omega^A_C\wedge\omega^C_B.
\end{equation}
Straightforward calculation gives for the torsion
\begin{eqnarray}
{\tt T^1}\!\!\!&=&\!\!\! d\theta^1+\omega^3\wedge\theta^2-\omega^2\wedge\theta^3-\sqrt{{2}}\,\omega^1\wedge(\theta^4+q\,\theta^5),\\
{\tt T^2}\!\!\!&=&\!\!\! d\theta^2+\omega^1\wedge\theta^3-\omega^3\wedge\theta^1-\sqrt{2}\bar q\,\omega^2(\theta^4+\theta^5),\\
{\tt T^3}\!\!\!&=&\!\!\! d\theta^3+\omega^2\wedge\theta^1-\omega^1\wedge\theta^2-\sqrt{{2}}\,\omega^3\wedge(q\,\theta^4+\theta^5),\\
{\tt T^4}\!\!\!&=&\!\!\! d\theta^4+\sqrt{{2}}(\omega^1\wedge\theta^1+q\,\omega^2\wedge\theta^2+\bar q\,\omega^3\wedge\theta^3),\\
{\tt T^5}\!\!\!&=&\!\!\! d\theta^5+\sqrt{{2}}(\bar q\,\omega^1\wedge\theta^1+q\,\omega^2\wedge\theta^2+\omega^3\wedge\theta^3),\\
\end{eqnarray}
and for the curvature
\begin{eqnarray}
{\tt R^1_2} = \zeta^{312},\;\;\;{\tt R^1_3} \!\!\!&=&\!\!\!-\zeta^{231},\;\;\;{\tt R^1_4} =-\sqrt{2}\,\zeta^{123},\;\;\;{\tt R^1_5}=-\sqrt{2}\, q\,\zeta^{123},\nonumber\\
{\tt R^2_3} \!\!\!&=&\!\!\! \zeta^{123},\;\;\;\;\;\;{\tt R^2_4}=-\sqrt{2}\,\bar q\,\zeta^{231},\;{\tt R^2_5}=-\sqrt{2}\,\bar q\,\zeta^{231},\nonumber\\
&&\qquad\quad\; {\tt R^3_4} = -\sqrt{2}\,\zeta^{312},\;\;\;{\tt R^3_5} = -\sqrt{2}\,\zeta^{312},\nonumber\\
&&\qquad\qquad\qquad\qquad\qquad\;\;\;\; {\tt R^4_5}=0,\nonumber
\end{eqnarray}
where $\zeta^{ijk}$ is a 2-form defined by $\zeta^{ijk}=d\omega^i+\omega^j\wedge\omega^k$, where $i,j,k$ is a cyclic permutation of integers $1,2,3$. It can be proved that a connection $\omega$ is consistent with a Hermitian metric $h$ and it preserves the tensor $K=(K_{AB})$, that is,
$$
\overset{\omega}{\nabla}\,h=0,\;\;\;\overset{w}{\nabla}\,K=0,
$$
where $\overset{\omega}\nabla$ is the covariant derivative of tensor fields induced by a connection $\omega$.
\section{Discussion}
In this paper, we study a ternary generalization of the concept of skew-symmetry, which is different from the classical one. This ternary analogue of skew-symmetry is defined by means of a faithful representation of the cyclic group $\mathbb Z_3$ by the cubic roots of unity $1,q,\bar q$. A ternary generalization of skew-symmetry considered in the present paper is motivated by the ternary generalization of the Pauli's exclusion principle proposed by R. Kerner in connection with quantum properties of quarks. The algebraic aspect of ternary generalization of skew-symmetry was studied in a number of scientific papers, where this generalization was used to construct ternary algebras with generators. Then these algebras were used to construct a generalization of the Dirac operator and of the calculus of differential forms. In this article we study a ternary generalization of the notion of skew-symmetry from the point of view of geometric structures. We consider the space of 3rd order covariant tensors, which are ternary skew-symmetric, that is, the sum of the tensor components obtained by cyclic permutations of subscripts is equal to zero. Moreover, the trace of a tensor over any pair of subscripts must be equal to zero. We think that this requirement is a ternary analogue of the fact that in the case of a 2nd order skew-symmetric (in the classical sense) covariant tensor, all diagonal elements (with equal subscripts) are equal to zero (and hence the sum, that is, the trace will be equal to zero). Tensors of 3rd order with the properties described above are known in the representation theory of the rotation group. They form a ten-dimensional space and in this space there is a twofold irreducible representation of the rotation group. In order to split this twofold representation into two irreducible representations we decompose this ten-dimensional space into a direct sum of two five-dimensional spaces with the help of the primitive cubic roots of unity $q=\exp(2\pi i/3), \bar q=\exp(4\pi i/3)$. This decomposition can be considered as some kind of duality, which is possibly related to the duality quark-anti-quark. We construct the five-dimensional complex Hermitian space whose points are identified with ternary skew-symmetric covariant  3rd order tensors. If we consider a 3rd order tensor as a 3-dimensional matrix, then we have a five-dimensional complex space whose points can be identified with 3-dimensional matrices. Figuratively speaking, we have a five-dimensional complex space, whose points are 3-dimensional lattices and the components of 3rd order ternary skew-symmetric covariant tensors are located at the nodes of these lattices. It is possible that a geometry of this five-dimensional complex Hermitian space is an appropriate geometric model for a space of our Universe at distances comparable with the sizes of quarks.




%


\end{document}